\documentclass[twocolumn,floats,showpacs,amsmath,amssymb,pre]{revtex4}

\usepackage{graphicx}

\begin{document}

\title{Solid helium at high pressure: A path-integral Monte Carlo simulation}
\author{Carlos P. Herrero}
\affiliation{Instituto de Ciencia de Materiales,
         Consejo Superior de Investigaciones Cient\'{\i}ficas (CSIC), 
         Campus de Cantoblanco, 28049 Madrid, Spain }
\date{\today}

\begin{abstract}
Solid helium ($^3$He and $^4$He) in the hcp and fcc phases has
been studied by path-integral Monte Carlo.
Simulations were carried out in the isothermal-isobaric ($NPT$) 
ensemble at pressures up to 52 GPa.
This allows one to study the temperature and pressure dependences of
isotopic effects on the crystal volume and vibrational energy in a
wide parameter range.
The obtained equation of state at room temperature agrees with available
experimental data.
The kinetic energy, $E_k$, of solid helium is found to be larger than 
the vibrational potential energy, $E_p$. The ratio $E_k/E_p$ amounts to 
about 1.4 at low pressures and decreases as the applied pressure is raised,
converging to 1, as in a harmonic solid.
Results of these simulations have been compared with those yielded by 
previous path integral simulations in the $NVT$ ensemble. 
The validity range of earlier approximations is discussed.  \\
\end{abstract}


\maketitle

\section{Introduction}

Structural and thermodynamic properties of solid helium have been of 
continuous interest
in condensed matter physics because of its quantum nature and electronic
simplicity. In fact, solid helium is in many respects an archetypal
``quantum solid,'' where zero-point energy and associated anharmonic 
effects are appreciably larger than in most known solids.
This gives rise to peculiar properties, whose understanding has presented 
a challence for elaborated theories and modelling from a microscopic
standpoint \cite{ce95}. 
Among these properties, the behavior of condensed helium at high density
has received much attention. 
In fact, diamond-anvil-cell and shock-wave experiments have allowed
to study the equation of state (EOS) of solid helium up to
pressures on the order of 50 GPa \cite{po86,ma88,lo93}.
In last years, the effect of pressure on heavier rare-gas solids
has also been of interest for both experimentalists \cite{sh01,er02} and
theorists \cite{ne00,ii01,de02,ts02}.

Anharmonic effects in solids, and in solid helium in particular,
have been traditionally studied by using theoretical techniques such
as quasiharmonic approximations and self-consistent phonon
theories \cite{kl76}.
In more recent years, the Feynman path-integral formulation of statistical
mechanics \cite{fe72,kl90} has been exploited to study thermodynamic 
properties of solids at temperatures lower than their Debye temperature
$\Theta_D$, where the quantum character of the atomic nuclei is relevant.
Monte Carlo sampling applied to evaluate finite-temperature
path integrals allows one to carry out quantitative and nonperturbative
studies of highly-anharmonic effects in solids \cite{ce95}.

 The path-integral Monte Carlo (PIMC) technique has been used to study 
several properties of solid helium \cite{ce95,ce96,ba89,bo94,ch01}, as well as 
heavier rare-gas solids \cite{cu93,mu95,ch02,ne02,he02}.
For helium, in particular, this method has predicted kinetic-energy 
values \cite{ce96} and Debye-Waller factors \cite{dr00} in good agreement with 
data derived from experiments \cite{ar03,ve03}.
PIMC simulations were also employed to study the isotopic shift
in the helium melting pressure \cite{ba89,bo94}.
The EOS of solid helium at T = 0 has been studied by
diffusion Monte Carlo in a wide density range (down to a molar volume of 15
cm$^3$/mol), using an accurate interatomic potential \cite{mo00}.
This has been done by Chang and Boninsegni \cite{ch01} at finite temperatures 
for both solid and liquid helium, by using PIMC simulations with several 
interatomic potentials, and for molar volumes down to 2.6 cm$^3$/mol.  
These authors suggested that the use
of effective potentials including two- and three-body terms alone can be
insufficient to reproduce the EOS of condensed helium, in the pressure
range experimentally accessible at present. 

In this paper, we study the effect of pressure on solid $^3$He and $^4$He 
by PIMC simulations. We employ the isothermal-isobaric ($NPT$) ensemble,
which allows us to consider properties of these solids along well-defined 
isobars.
The interatomic interaction is described by a combination of two- and
three-body terms, which are directly included in the simulations. This
permits us to check results of earlier simulations, where the effect of 
three-body terms was taken into account in a perturbative way \cite{bo94,ch01}.
By comparing results for $^3$He and $^4$He, we analyse isotopic
effects on the vibrational energy and crystal volume of solid helium.

The paper is organized as follows.  In Sec.\,II, the
computational method is described. In Sec.\,III we present results for 
the equation of state, vibrational energy, and isotopic effects
on the lattice parameters. Finally, Sec.\,IV includes a discussion of 
the results and the conclusions.

\section{Method}
                                                                                    
Equilibrium properties of solid helium in the face-centred cubic (fcc) and
hexagonal close-packed (hcp) phases have been
calculated by PIMC simulations in the $NPT$ ensemble.
Most of our simulations were performed on supercells of the fcc and hcp
unit cells, including 500 and 432 helium atoms respectively.
To check the convergence of our results with system size, some simulations
were carried out for other supercell sizes, and it was found that
finite-size effects for $N > 400$ atoms are negligible for the quantities
studied here (they are smaller than the error bars).

Helium atoms have been treated as quantum particles interacting
through an effective interatomic potential, composed of a two-body
and a three-body part.  For the two-body interaction, we have
taken the potential developed by Aziz {\em et al.} \cite{az95}
(the so-called HFD-B3-FCI1 potential). For the three-body
part we have employed a Bruch-McGee-type potential \cite{br73,lo87},
which includes the exchange three-body interaction and a triple-dipole
Axilrod-Teller interaction. 
For most of the simulations presented below, the parameters employed for
the three-body terms were those given by Loubeyre \cite{lo87}, but with
the parameter $A$ in the attractive exchange interaction rescaled by a 
factor 2/3 (as suggested in Ref. \cite{bo94}, and giving $A = 20.43$ au).
                                                                                    
In the path-integral formulation of statistical mechanics, the partition
function is evaluated through a discretization of the density matrix
along cyclic paths, composed of a finite number $L$ (Trotter number)
of ``imaginary-time'' steps \cite{fe72,kl90}. In the numerical simulations,
this discretization gives rise to the appearance of $L$ replicas
for each quantum particle.  Thus, the practical implementation of this 
method relies on an isomorphism between the quantum system and a classical 
one, obtained by replacing each quantum particle 
by a cyclic chain of $L$ classical particles, connected
by harmonic springs with a temperature-dependent constant.
Details on this computational method can be found
elsewhere \cite{gi88,ce95,no96}.

Our simulations were based on the so-called ``primitive'' form
of PIMC \cite{ch81,si88}.
For interatomic potentials including only two-body terms, effective
forms for the density matrix have been developed, which allow one to
reduce efficiently the Trotter number, thereby simplifying appreciably the
calculation \cite{bo94}.
Such a simplification is not possible here, since we consider explicitly
three-body terms in the simulations.
Quantum exchange effects between atomic nuclei were not considered,
as they are negligible for
solid helium at the pressures and temperatures studied here. (This should be
valid as long as there are no vacancies and $T$ is greater than the
exchange frequency $\sim 10^{-6}$ K \cite{ce95}.)
The dynamic effect of the interactions between nearest
and next-nearest neighbours is explicitly considered.  The effect
of interactions beyond next-nearest neighbours is taken into
account by a static-lattice approximation \cite{cu97,mu95},
which was employed earlier in PIMC simulations of
rare-gas solids \cite{ne00,he02}.
We have checked that including dynamical correlations between more
distant atom shells does not change the results presented below.
For the energy we have used the ``crude'' estimator, as defined in 
Refs. \cite{ch81,si88}.
All calculations have been performed using our PIMC code \cite{no96b},
that has been employed earlier to study various types of 
solids \cite{no96,he01,he02,he05}.

Sampling of the configuration space has been carried out by the Metropolis
method at pressures $P \leq$ 52 GPa, and temperatures between 25 K and the
melting temperature of the solid at each considered pressure.
A simulation run proceeds via successive Monte Carlo steps.
In each step, the replica coordinates are updated according to three 
different kinds of sampling schemes: (1) Sequential trial moves of the 
individual replica coordinates; (2) trial moves of the center of gravity 
of the cyclic paths, keeping unaltered the shape of each path, and 
(3) trial changes on the logarithm of the volume of the simulation cell. 
 For given temperature and pressure, a typical run consisted
of $10^4$ Monte Carlo steps for system equilibration, followed by $10^5$ 
steps for the calculation of ensemble average properties.
Other technical details are the same as those used in Refs.
\cite{he02,he03a}.

To have a nearly constant precision for the simulation results
at different temperatures, we have taken a Trotter number that
scales as the inverse temperature. At a given $T$, the 
value of $L$ required to obtain convergence of the results depends
on the Debye temperature $\Theta_D$ of the considered solid 
(higher $\Theta_D$ needs larger $L$). 
Since vibrational frequencies (and the associated Debye temperature) 
increase as the applied pressure is raised, $L$ has to be increased 
accordingly.  Thus, we have taken $L T$ = 4000 K for $^3$He and 3000 K for 
$^4$He, which were found to be sufficient for simulations of the corresponding
solids at the pressures considered here ($P \leq$ 52 GPa). 
This means that, for solid $^3$He at $T =$ 25 K, we have $L$ = 160
and then the computational time required to carry out a PIMC simulation 
for $N =500$ helium atoms is equivalent to a classical Monte Carlo simulation 
of $L N$ = 80000 atoms (assuming the same number of simulation steps).

\section{Results}

\subsection{Crystal volume}

\begin{figure}
\vspace{-2cm}
\hspace*{-0.8cm}
\includegraphics[width=9.5cm]{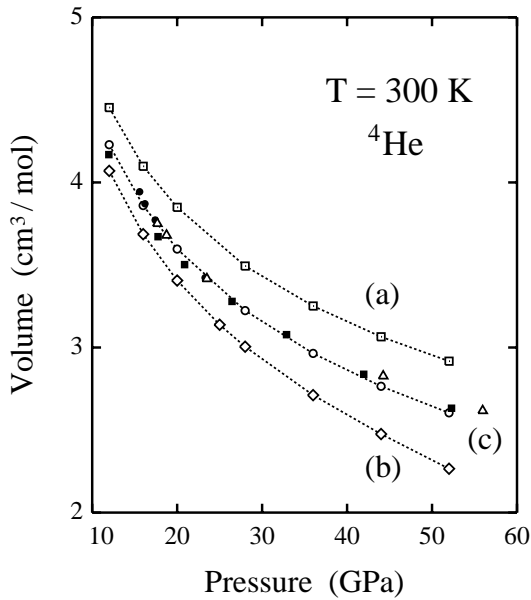}
\vspace{-2.6cm}
\caption{\label{f1}
Equation of state pressure-volume of hcp $^4$He at
300 K, as derived from PIMC simulations for different interatomic
potentials. (a) Open squares: only two-body interactions with an Aziz-type
potential \cite{az95}; (b) open diamonds: two-body terms as in
Ref. \cite{az95} and three-body interactions, as defined in Ref. \cite{lo87};
(c) open circles: same potential as (b), but with the exchange three-body
interaction rescaled by 2/3.
Error bars of the simulation results are less than the symbol size.
 Dotted lines are guides to the eye.
Open triangles show earlier results obtained from PIMC simulations in the
$NVT$ ensemble \cite{ch01}, with the attractive exchange interaction
rescaled as in (c).  Black symbols show experimental data obtained by Mao
{\em et al.} \cite{ma88} (filled circles) and Loubeyre {\em et al.}
\cite{lo93} (filled squares).
}
\end{figure}

Shown in Fig. 1 is the pressure dependence of the crystal volume for
solid $^4$He at 300 K. Open symbols represent results of PIMC simulations
with different interatomic potentials: (a) only two-body interactions
with an Aziz-type potential \cite{az95} (squares); (b) two-body interactions
as in Ref. \cite{az95} plus three-body terms as in Ref. \cite{lo87} 
(diamonds);
(c) the same two-body potential and three-body interaction with the exchange
part rescaled by 2/3, as proposed in Ref. \cite{bo94} (circles).
For comparison, we also present results derived from PIMC simulations in
the $NVT$ ensemble \cite{ch01}, with the exchange interaction rescaled by the 
same factor 2/3 (triangles).
Filled symbols indicate experimental results obtained by Mao 
{\em et al.} \cite{ma88} (filled circles) and Loubeyre {\em et al.} \cite{lo93}
(filled squares).
Our simulation data show that the interatomic potential (c) gives results
for the equation of state of solid $^4$He close to the experimental data.
The only consideration of two-body terms predicts, for a given applied
pressure, a crystal volume larger than the experimental one. On the
contrary, consideration of both two- and three-body terms derived from
{\em ab initio} calculations underestimates the volume of solid helium. 
This is in line with results obtained earlier from PIMC simulations 
in the $NVT$ ensemble in Refs. \cite{bo94,ch01}.
These authors introduced the three-body terms not directly in the simulations, 
but calculated in a perturbative way their contribution to thermodynamic
averages from configurations obtained in PIMC simulations. 
In  Fig. 1 (open circles and triangles) we observe that our $NPT$
simulations give the same results as those obtained earlier in the
$NVT$ ensemble \cite{ch01} for pressures lower than 30 GPa. However, at higher
pressures both sets of results differ appreciably, and for
a given $P$ the procedure employed in Ref. \cite{ch01}
yields a volume larger than that obtained here.

Our results with the effective interatomic potential (c) follow closely the 
experimental ones, even at the highest pressures considered here, although they 
seem to become lower than the later as pressure rises.  
However, taking into account the dispersion of experimental points and the 
error bars of the simulation data, differences between both sets of data
are not enough to invalidate the accuracy of the effective 
potential (c) in the pressure range studied here. 
This contrasts with the conclusions presented
by Chang and Boninsegni \cite{ch01}, who argued that two- and three-body
terms alone may be insufficent to reproduce quantitatively the EOS of 
condensed helium at pressures on the order of 50 GPa.
In fact, for molar volumes smaller than 3 cm$^3$/mol, these authors found 
pressures larger than the experimental ones (see Fig. 1).
In view of these results, in the remainder of the paper we will employ only 
the interaction potential (c) (with the exchange three-body part of 
Ref. \cite{lo87} rescaled by 2/3).

\begin{figure}
\vspace{0.2cm}
\hspace*{-2.0cm}
\includegraphics[width=13cm]{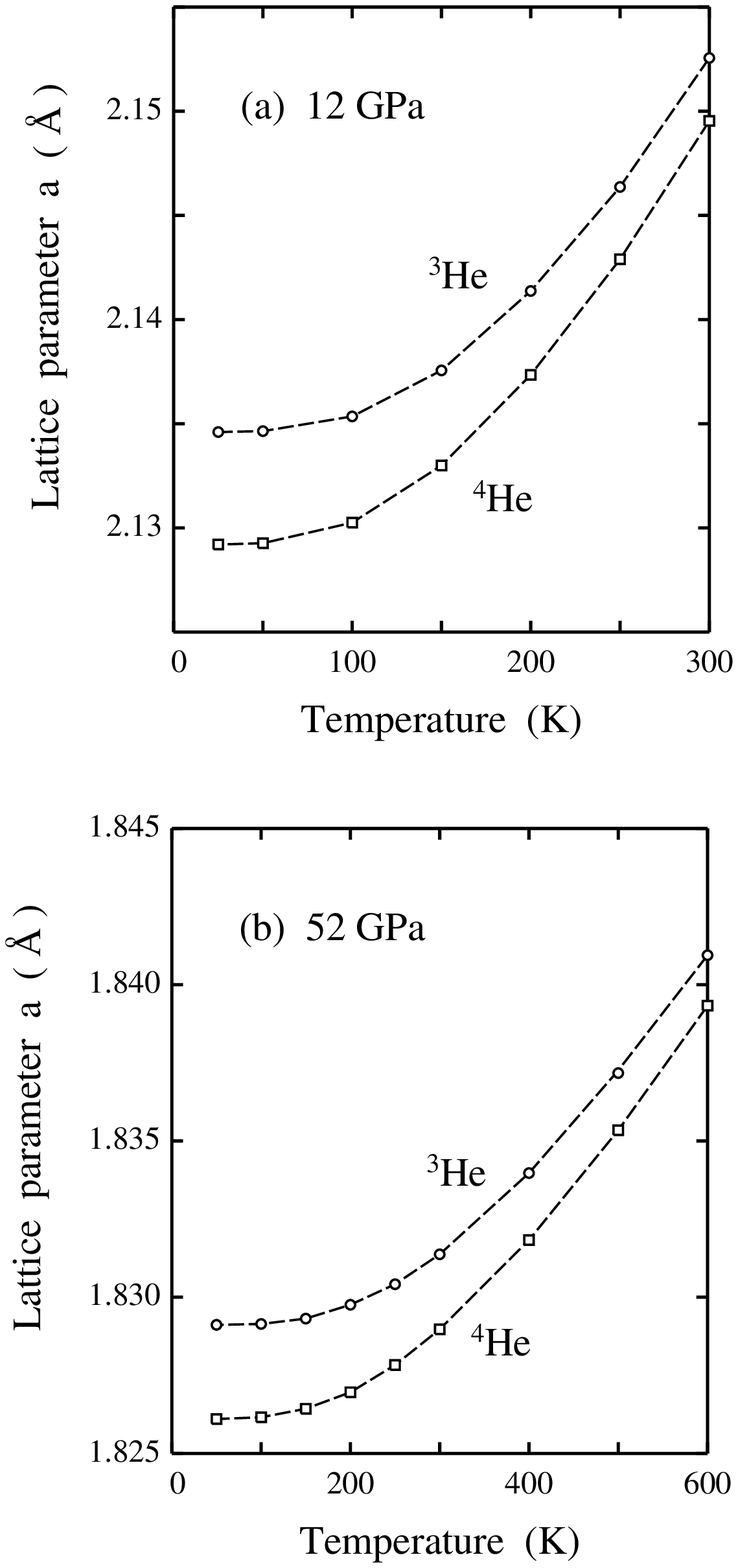}
\vspace{-1.1cm}
\caption{\label{f2}
Temperature dependence of the lattice parameter $a$ of hcp helium, derived
from PIMC simulations at two different pressures: (a) 12 GPa and (b) 52 GPa.
Squares and circles indicate results for $^4$He and $^3$He, respectively.
Error bars are smaller than the symbol size.
Dashed lines are guides to the eye.
}
\end{figure}

In Fig. 2 we present the temperature dependence of the lattice parameter
$a$ for hcp helium at two pressures: (a) 12 GPa and (b) 52 GPa. 
Squares and circles represent results of our PIMC simulations for
$^4$He and $^3$He, respectively.
Note the different vertical and horizontal scales in (a) and (b).
As expected, the difference $a_3 - a_4$ between lattice parameters of 
$^3$He and $^4$He at a given pressure decreases as the temperature is raised 
(at high $T$ the solid becomes ``more classical").

\begin{figure}
\vspace{-2cm}
\hspace*{-0.8cm}
\includegraphics[width=9.5cm]{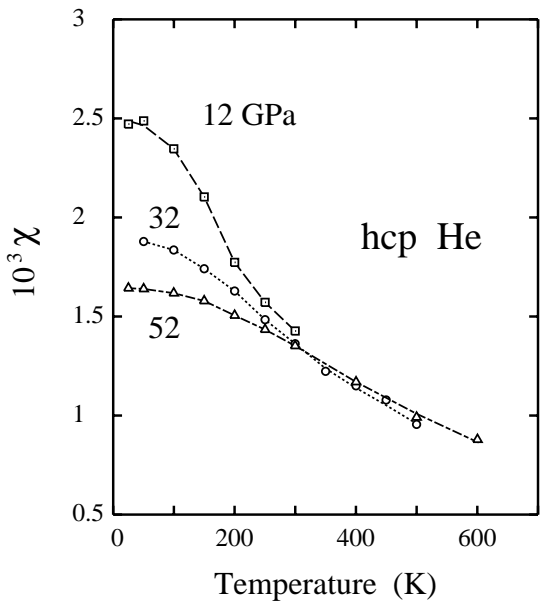}
\vspace{-2.6cm}
\caption{\label{f3}
Isotopic effect on the lattice parameter $a$ of hcp helium, as obtained from
PIMC simulations. Shown is the
parameter $\chi = (a_3 - a_4) / a_4$ as a function of temperature for
three different pressures: squares, 12 GPa; circles, 32 GPa; triangles,
52 GPa. Error bars are on the order of the symbol size. Lines are guides
to the eye.
}
\end{figure}

To quantify the isotopic effect on the linear dimensions of solid helium,
we employ the parameter $\chi = (a_3 - a_4) / a_4$,
which measures the relative difference between lattice parameters of
solid $^3$He and $^4$He. This parameter $\chi$ is displayed in Fig. 3
as a function of temperature for three different pressures:
12 (squares), 32 (circles), and 52 GPa (triangles). For each pressure,
results are shown for temperatures at which the considered solids were
stable along the PIMC simulations.
At low $T$, the parameter $\chi$ is larger for lower pressure, and
values obtained for different pressures approach each other as the
temperature rises. At $T > 300$ K, we find that $\chi$
is larger for $P$ = 52 GPa than for 32 GPa. This does not mean
that the difference $a_3 - a_4$ is larger at 52 GPa, but is a consequence
of the normalization by $a_4$, which is clearly smaller at higher pressure.

The difference $a_3 - a_4$ is largest at small pressures and low
temperatures, where quantum effects are most prominent. 
For a given solid, quantum effects on the crystal size can be measured
by the difference $\Delta a = a - a_{\rm cl}$
between the actual lattice parameter $a$ and that obtained for
a ``classical'' crystal of point particles, $a_{\rm cl}$.
This difference decreases for increasing atomic mass and 
temperature \cite{mu95,he01}.
From our PIMC simulations at $T$ = 25 K and a relatively low pressure
of 0.3 GPa, we found an increase in the linear size of solid $^3$He and
$^4$He of 8.2 and 7.2\% with respect to the classical crystal at
zero temperature.
For comparison, we note that the ``zero-point expansion" for
heavier rare gases causes a relative increase of 4.1\% and 1.2\% in the 
lattice parameter of fcc Ne and Ar, respectively \cite{he03a}.

\subsection{Energy}

For a given interatomic potential, the internal energy of a solid,
$E(V,T)$, at volume $V$ and temperature $T$ can be written as:
\begin{equation}
  E(V,T) =  E_{\rm min}(V) + E_{\rm vib}(V,T)   \, ,
\label{evt}
\end{equation}
where $E_{\rm min}(V)$ is the potential energy for the (classical) crystal 
at $T = 0$ with point atoms on their lattice sites, and $E_{\rm vib}(V,T)$
is the vibrational energy.
Since we are working here in the isothermal-isobaric ensemble, the volume 
is implicitly given by the applied pressure.
At finite temperatures, $V$ changes with $T$ due to thermal expansion,
and for real (quantum) solids, the crystal volume depends on 
quantum effects, which also contribute to expand the crystal with respect to
the classical expectancy (mainly at low $T$).
Thus, the volume $V$ in Eq. (\ref{evt}) is an implicit function of $P$ and
$T$, i.e., $V = V(P,T)$.

The vibrational energy, $E_{\rm vib}(V,T)$, depends explicitly on both,
$V$ and $T$, and can be obtained by subtracting the energy $E_{\rm min}(V)$
from the internal energy.
In this way, path-integral Monte Carlo simulations allow one to obtain
separately the kinetic, $E_k$, and potential energy, $E_p$,
associated to the lattice vibrations:  $E_{\rm vib} = E_k + E_p$.

\begin{figure}
\vspace{-2cm}
\hspace*{-0.4cm}
\includegraphics[width=9.5cm]{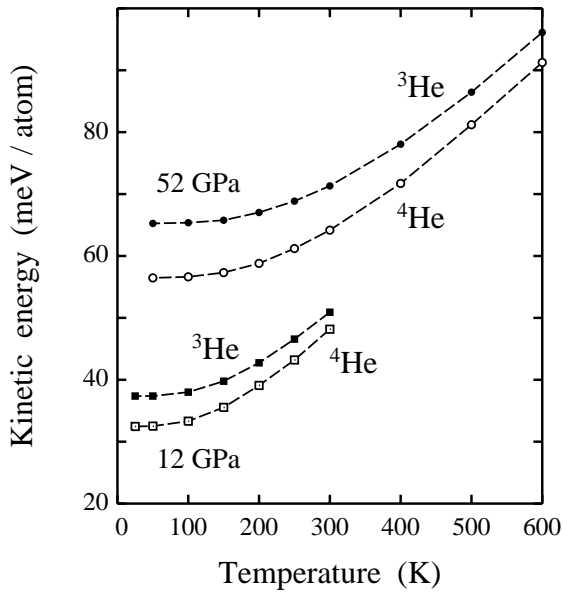}
\vspace{-2.6cm}
\caption{\label{f4}
Temperature dependence of the kinetic energy of hcp solid helium at two
different pressures, as derived from PIMC simulations: $P$ = 12 GPa
(squares) and 52 GPa (circles).
Open and filled symbols correspond to $^4$He and $^3$He, respectively.
Error bars of the simulation results are less than the symbol size.
 Dashed lines are guides to the eye.
}
\end{figure}

In Fig. 4 we show the kinetic energy of solid helium (hcp phase) as 
a function of temperature for $P$ = 12 and 52 GPa. Open symbols correspond
to $^4$He and filled symbols to $^3$He.
At 52 GPa, the low-temperature kinetic energy of solid $^4$He is found to
be 56 meV/atom, to be compared with 8.3 meV/atom obtained for 
the fcc phase at 25 K and low pressure ($P$ = 0.3 GPa, giving a molar
volume of 9.95 cm$^3$/mol).
This value of $E_k$ is similar to those obtained earlier from
PIMC simulations in the $NVT$ ensemble at temperatures close to 25 K
and molar volumes around 10 cm$^3$/mol \cite{dr00}. 
 
At low temperature the ratio $E_k^3 / E_k^4$ between the kinetic energy 
of $^3$He and $^4$He is close to 1.155, 
the expected value in a harmonic model of lattice
vibrations. This indicates that, irrespective of the important anharmonicity 
present in these solids, anharmonic shifts in $E_k$ scale with
mass approximately as in a quasiharmonic model.
The ratio $E_k^3 / E_k^4$ decreases as temperature rises and quantum effects
are less important. At the highest simulated temperatures (300 K for 
$P$ = 12 GPa and 600 K for 52 GPa) we find $E_k^3 / E_k^4 \approx 1.05$,
still clearly larger than the classical limit: $E_k^3 / E_k^4 \to 1$.

\begin{figure}
\vspace{-2cm}
\hspace*{-0.8cm}
\includegraphics[width=9.5cm]{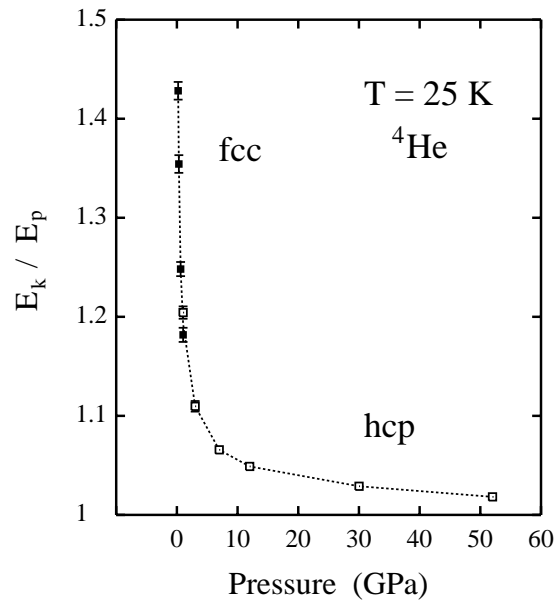}
\vspace{-2.6cm}
\caption{\label{f5}
Kinetic-to-potential energy ratio, $E_k / E_p$, for solid $^4$He as a
function of applied pressure at $T$ = 25 K.
Black and open symbols correspond to fcc and hcp phases, respectively.
Error bars, where not shown, are smaller than the symbol size.
}
\end{figure}

The anharmonicity of the lattice vibrations is clearly noticeable 
when one compares the kinetic and potential energy of solid helium. 
From our PIMC simulations, we have found in all cases considered here
that the vibrational potential energy $E_p$ is smaller than the kinetic 
energy $E_k$.
Shown in Fig. 5 is the ratio $E_k / E_p$ for $^4$He as a function of
pressure at 25 K. We have included in this figure results for fcc (low
pressure) and hcp helium (high pressure). The kinetic-to-potential energy
ratio decreases with increasing pressure; first it gows down fast for
pressures lower than 5 GPa, and then it decreases slower at higher
pressures, approaching the value  $E_k / E_p$ = 1, characteristic of 
harmonic vibrations.
We note that the fact $E_k > E_p$ has been also found for solids of 
heavier rare gases from PIMC simulations \cite{mu95,he02,he03a}.

For solid $^4$He at low pressure, we find that $E_k$ is larger than $E_p$ 
by about 40\%.
This important difference between $E_k$ and $E_p$ reflects the large
anharmonicity of lattice vibrations in this solid. For comparison,
we mention that these energies differ by about 20\% and 7\% for solid
neon and argon, respectively \cite{he02,he03a}. These numbers are very 
large when compared with covalent or metallic solids, and are due to the 
weakness of van der Waals-type bonds, which allows for a large amplitude of 
the atomic vibrations.

\section{Discussion}

Path-integral Monte Carlo simulations are well suited 
to study quantum effects on structural and thermodynamic properties of
solids. These effects are particularly important for solid helium, where
isotopic effects are relevant, as manifested in differences between 
crystal volume and vibrational energies of solid $^3$He and $^4$He.
The PIMC method enables us to study phonon-related properties without
the assumptions of quasiharmonic or self-consistent phonon approximations, 
and to study anharmonic effects in solids in a nonperturbative way.
This method allows us to separate the kinetic and potential
contributions to the vibrational energy of a given solid, and
quantify the anharmonicity of the lattice vibrations, which
together with zero-point motion give rise to isotopic effects. 

PIMC simulations yield in principle ``exact'' values for measurable
properties of many-body quantum problems, with an accuracy limited by the 
imaginary-time step (Trotter number) and the statistical error of the Monte 
Carlo sampling.  Thus, the effectve interatomic potential employed here gives 
a good description of structural and thermodynamic properties of solid
helium up to pressures on the order of those presently achieved in
experiments.
In particular, the equation of state $P$-$V$ is well described by this 
potential in the pressure range considered here ($P \lesssim 50$ GPa),
even though it is partially an empirical potential, in the sense that the
attractive three-body part was rescaled {\em ad hoc} from 
the fit to {\em ab initio} calculations. 
Explicit consideration of three-body terms in PIMC simulations makes 
the computation times much longer, in comparison with simulations
including only two-body terms. But this has allowed us to check the
validity limit of some approximations made earlier in this kind of 
calculations, such as including the effect of three-body terms in
a perturbative way (from configurations obtained in PIMC 
simulations) \cite{bo94,ch01}.
As a result, this approximation yields at $P < 30$ GPa an EOS 
indistinguishable within error bars from that obtained here.
At higher pressures, however, it predicts volumes larger than those 
found in the present work (and in experiment).  

Due to the large anharmonicity of lattice vibrations in
solid helium, the isotopic effect on the crystal volume
is important. 
At low temperatures, the dependence of the crystal volume
on isotopic mass is related to the zero-point lattice expansion.
The magnitude of this isotopic effect decreasees appreciably as temperature
or pressure are raised.  Nevertheless, we emphasize that it is still
measurable at the highest pressures considered here and at $T$ close
to the corresponding melting temperature, as shown in Fig. 3. 
This is in line with the observed 
isotopic effect on the melting temperature at low pressure \cite{ba89,bo94}.

Anharmonic effects in solids have been quantified earlier
by comparing the kinetic and potential energies derived from
PIMC simulations \cite{mu95,he03a}.
For solid helium, and for different pressures and
temperatures, we have found $E_k > E_p$, as for solids of heavier
rare gases with Lennard-Jones-type potentials \cite{he03a}. 
The departure of the ratio $E_k / E_p$ from its value for a purely
harmonic solid ($E_k / E_p = 1$) is a measure of the overall anharmonicity
of the considered solids.
For given $T$, we have found that this energy ratio decreases as pressure
is raised.
Note that the ratio $E_k/E_p$ alone does not give any information
on the details of the anharmonicity of the vibrational modes, but gives a 
quantitative measure of the overall anharmonicity of a given solid.
In particular, it is well suited to compare the anharmonicity of similar
materials. For rare-gas solids at low $T$, $E_k/E_p$ increases from
1.02 for xenon to 1.23 for neon at zero pressure \cite{he03a}. This ratio is 
clearly higher for solid helium at low $P$, as shown in Fig. 5:
$E_k/E_p \approx$ 1.4 for fcc $^4$He at 25 K. 

It has been recently suggested that pressure causes a decrease in 
anharmonicity \cite{ka03,la04,he05}, in line with earlier observations that
the accuracy of quasiharmonic approximations increases as pressure rises 
and the density of the material increases \cite{po72,ho73}. 
Thus, such approximations become exact in the high-density
limit \cite{ho73,he05}.
This is related to the fact that the ratio of the vibrational energy to 
the whole internal energy of the crystal decreases for increasing pressure,
in spite of the increase in zero-point energy caused by the rise in
vibrational frequencies.
It has been also argued that at high pressures, thermodynamic properties
of solids can be well described by classical calculations, i.e., dealing
with the atoms as classical oscillators in a given potential \cite{et74}.
This means that, at a given temperature, such a classical approach becomes 
more and more accurate as pressure is raised.
The origin of this is similar to that described above for the success of 
quasiharmonic approaches. Since in this respect the lattice vibrations become 
less relevant as pressure rises, and eventually give a negligible contribution
to the free energy of the solid, their description by a classical
or a quantum model becomes unimportant for solids in the limit of very 
large pressures.

In summary, we have carried out PIMC simulations of solid helium in the
isothermal-isobaric ensemble.
In general, our results support the assumptions made in earlier
calculations, where three-body terms were included in a perturbative
way. In particular, the equation of state obtained here coincides at
$P \lesssim 30$ GPa with 
that found in calculations employing the $NVT$ ensemble with that assumption.
At high pressures, however, data yielded by the present PIMC simulations 
differ from those obtained earlier, using similar interatomic potentials.
 The kind of effective interatomic potentials 
employed here, including two- and three-body terms will probably give
a poor description of solid helium at very high pressures 
($\gtrsim$ 60 GPa), but seems appropriate for the pressure range in
which experimental data are available at present.

\begin{acknowledgments}
The author benefitted from discussions with R. Ram\'{\i}rez.
This work was supported by CICYT (Spain) through Grant
No. BFM2003-03372-C03-03.   \\
\end{acknowledgments}

\bibliographystyle{apsrev}

\begin{thebibliography}{43}
\expandafter\ifx\csname natexlab\endcsname\relax\def\natexlab#1{#1}\fi
\expandafter\ifx\csname bibnamefont\endcsname\relax
  \def\bibnamefont#1{#1}\fi
\expandafter\ifx\csname bibfnamefont\endcsname\relax
  \def\bibfnamefont#1{#1}\fi
\expandafter\ifx\csname citenamefont\endcsname\relax
  \def\citenamefont#1{#1}\fi
\expandafter\ifx\csname url\endcsname\relax
  \def\url#1{\texttt{#1}}\fi
\expandafter\ifx\csname urlprefix\endcsname\relax\def\urlprefix{URL }\fi
\providecommand{\bibinfo}[2]{#2}
\providecommand{\eprint}[2][]{\url{#2}}

\bibitem[{\citenamefont{Ceperley}(1995)}]{ce95}
\bibinfo{author}{\bibfnamefont{D.~M.} \bibnamefont{Ceperley}},
  \bibinfo{journal}{Rev. Mod. Phys.} \textbf{\bibinfo{volume}{67}},
  \bibinfo{pages}{279} (\bibinfo{year}{1995}).

\bibitem[{\citenamefont{Polian and Grimsditch}(1986)}]{po86}
\bibinfo{author}{\bibfnamefont{A.}~\bibnamefont{Polian}} \bibnamefont{and}
  \bibinfo{author}{\bibfnamefont{M.}~\bibnamefont{Grimsditch}},
  \bibinfo{journal}{Europhys. Lett.} \textbf{\bibinfo{volume}{2}},
  \bibinfo{pages}{849} (\bibinfo{year}{1986}).

\bibitem[{\citenamefont{Mao et~al.}(1988)\citenamefont{Mao, Hemley, Wu,
  Jephcoat, Finger, Zha, and Bassett}}]{ma88}
\bibinfo{author}{\bibfnamefont{H.~K.} \bibnamefont{Mao}},
  \bibinfo{author}{\bibfnamefont{R.~J.} \bibnamefont{Hemley}},
  \bibinfo{author}{\bibfnamefont{Y.}~\bibnamefont{Wu}},
  \bibinfo{author}{\bibfnamefont{A.~P.} \bibnamefont{Jephcoat}},
  \bibinfo{author}{\bibfnamefont{L.~W.} \bibnamefont{Finger}},
  \bibinfo{author}{\bibfnamefont{C.~S.} \bibnamefont{Zha}}, \bibnamefont{and}
  \bibinfo{author}{\bibfnamefont{W.~A.} \bibnamefont{Bassett}},
  \bibinfo{journal}{Phys. Rev. Lett.} \textbf{\bibinfo{volume}{60}},
  \bibinfo{pages}{2649} (\bibinfo{year}{1988}).

\bibitem[{\citenamefont{Loubeyre et~al.}(1993)\citenamefont{Loubeyre,
  LeToullec, Pinceaux, Mao, Hu, and Hemley}}]{lo93}
\bibinfo{author}{\bibfnamefont{P.}~\bibnamefont{Loubeyre}},
  \bibinfo{author}{\bibfnamefont{R.}~\bibnamefont{LeToullec}},
  \bibinfo{author}{\bibfnamefont{J.~P.} \bibnamefont{Pinceaux}},
  \bibinfo{author}{\bibfnamefont{H.~K.} \bibnamefont{Mao}},
  \bibinfo{author}{\bibfnamefont{J.}~\bibnamefont{Hu}}, \bibnamefont{and}
  \bibinfo{author}{\bibfnamefont{R.~J.} \bibnamefont{Hemley}},
  \bibinfo{journal}{Phys. Rev. Lett.} \textbf{\bibinfo{volume}{71}},
  \bibinfo{pages}{2272} (\bibinfo{year}{1993}).

\bibitem[{\citenamefont{Shimizu et~al.}(2001)\citenamefont{Shimizu, Tashiro,
  Kume, and Sasaki}}]{sh01}
\bibinfo{author}{\bibfnamefont{H.}~\bibnamefont{Shimizu}},
  \bibinfo{author}{\bibfnamefont{H.}~\bibnamefont{Tashiro}},
  \bibinfo{author}{\bibfnamefont{T.}~\bibnamefont{Kume}}, \bibnamefont{and}
  \bibinfo{author}{\bibfnamefont{S.}~\bibnamefont{Sasaki}},
  \bibinfo{journal}{Phys. Rev. Lett.} \textbf{\bibinfo{volume}{86}},
  \bibinfo{pages}{4568} (\bibinfo{year}{2001}).

\bibitem[{\citenamefont{Errandonea et~al.}(2002)\citenamefont{Errandonea,
  Schwager, Boehler, and Ross}}]{er02}
\bibinfo{author}{\bibfnamefont{D.}~\bibnamefont{Errandonea}},
  \bibinfo{author}{\bibfnamefont{B.}~\bibnamefont{Schwager}},
  \bibinfo{author}{\bibfnamefont{R.}~\bibnamefont{Boehler}}, \bibnamefont{and}
  \bibinfo{author}{\bibfnamefont{M.}~\bibnamefont{Ross}},
  \bibinfo{journal}{Phys. Rev. B} \textbf{\bibinfo{volume}{65}},
  \bibinfo{pages}{214110} (\bibinfo{year}{2002}).

\bibitem[{\citenamefont{Neumann and Zoppi}(2000)}]{ne00}
\bibinfo{author}{\bibfnamefont{M.}~\bibnamefont{Neumann}} \bibnamefont{and}
  \bibinfo{author}{\bibfnamefont{M.}~\bibnamefont{Zoppi}},
  \bibinfo{journal}{Phys. Rev. B} \textbf{\bibinfo{volume}{62}},
  \bibinfo{pages}{41} (\bibinfo{year}{2000}).

\bibitem[{\citenamefont{Iitaka and Ebisuzaki}(2001)}]{ii01}
\bibinfo{author}{\bibfnamefont{T.}~\bibnamefont{Iitaka}} \bibnamefont{and}
  \bibinfo{author}{\bibfnamefont{T.}~\bibnamefont{Ebisuzaki}},
  \bibinfo{journal}{Phys. Rev. B} \textbf{\bibinfo{volume}{65}},
  \bibinfo{pages}{012103} (\bibinfo{year}{2001}).

\bibitem[{\citenamefont{Dewhurst et~al.}(2002)\citenamefont{Dewhurst, Ahuja,
  Li, and Johansson}}]{de02}
\bibinfo{author}{\bibfnamefont{J.~K.} \bibnamefont{Dewhurst}},
  \bibinfo{author}{\bibfnamefont{R.}~\bibnamefont{Ahuja}},
  \bibinfo{author}{\bibfnamefont{S.}~\bibnamefont{Li}}, \bibnamefont{and}
  \bibinfo{author}{\bibfnamefont{B.}~\bibnamefont{Johansson}},
  \bibinfo{journal}{Phys. Rev. Lett.} \textbf{\bibinfo{volume}{88}},
  \bibinfo{pages}{075504} (\bibinfo{year}{2002}).

\bibitem[{\citenamefont{Tsuchiya and Kawamura}(2002)}]{ts02}
\bibinfo{author}{\bibfnamefont{T.}~\bibnamefont{Tsuchiya}} \bibnamefont{and}
  \bibinfo{author}{\bibfnamefont{K.}~\bibnamefont{Kawamura}},
  \bibinfo{journal}{J. Chem. Phys.} \textbf{\bibinfo{volume}{117}},
  \bibinfo{pages}{5859} (\bibinfo{year}{2002}).

\bibitem[{\citenamefont{Klein and Venables}(1976)}]{kl76}
\bibinfo{editor}{\bibfnamefont{M.~L.} \bibnamefont{Klein}} \bibnamefont{and}
  \bibinfo{editor}{\bibfnamefont{J.~A.} \bibnamefont{Venables}}, eds.,
  \emph{\bibinfo{title}{Rare Gas Solids}} (\bibinfo{publisher}{Academic Press},
  \bibinfo{address}{New York}, \bibinfo{year}{1976}).

\bibitem[{\citenamefont{Feynman}(1972)}]{fe72}
\bibinfo{author}{\bibfnamefont{R.~P.} \bibnamefont{Feynman}},
  \emph{\bibinfo{title}{Statistical Mechanics}}
  (\bibinfo{publisher}{Addison-Wesley}, \bibinfo{address}{New York},
  \bibinfo{year}{1972}).

\bibitem[{\citenamefont{Kleinert}(1990)}]{kl90}
\bibinfo{author}{\bibfnamefont{H.}~\bibnamefont{Kleinert}},
  \emph{\bibinfo{title}{Path Integrals in Quantum Mechanics, Statistics and
  Polymer Physics}} (\bibinfo{publisher}{World Scientific},
  \bibinfo{address}{Singapore}, \bibinfo{year}{1990}).

\bibitem[{\citenamefont{Ceperley et~al.}(1996)\citenamefont{Ceperley, Simmons,
  and Blasdell}}]{ce96}
\bibinfo{author}{\bibfnamefont{D.~M.} \bibnamefont{Ceperley}},
  \bibinfo{author}{\bibfnamefont{R.~O.} \bibnamefont{Simmons}},
  \bibnamefont{and} \bibinfo{author}{\bibfnamefont{R.~C.}
  \bibnamefont{Blasdell}}, \bibinfo{journal}{Phys. Rev. Lett.}
  \textbf{\bibinfo{volume}{77}}, \bibinfo{pages}{115} (\bibinfo{year}{1996}).

\bibitem[{\citenamefont{Barrat et~al.}(1989)\citenamefont{Barrat, Loubeyre, and
  Klein}}]{ba89}
\bibinfo{author}{\bibfnamefont{J.~L.} \bibnamefont{Barrat}},
  \bibinfo{author}{\bibfnamefont{P.}~\bibnamefont{Loubeyre}}, \bibnamefont{and}
  \bibinfo{author}{\bibfnamefont{M.~L.} \bibnamefont{Klein}},
  \bibinfo{journal}{J. Chem. Phys.} \textbf{\bibinfo{volume}{90}},
  \bibinfo{pages}{5644} (\bibinfo{year}{1989}).

\bibitem[{\citenamefont{Boninsegni et~al.}(1994)\citenamefont{Boninsegni,
  Pierleoni, and Ceperley}}]{bo94}
\bibinfo{author}{\bibfnamefont{M.}~\bibnamefont{Boninsegni}},
  \bibinfo{author}{\bibfnamefont{C.}~\bibnamefont{Pierleoni}},
  \bibnamefont{and} \bibinfo{author}{\bibfnamefont{D.~M.}
  \bibnamefont{Ceperley}}, \bibinfo{journal}{Phys. Rev. Lett.}
  \textbf{\bibinfo{volume}{72}}, \bibinfo{pages}{1854} (\bibinfo{year}{1994}).

\bibitem[{\citenamefont{Chang and Boninsegni}(2001)}]{ch01}
\bibinfo{author}{\bibfnamefont{S.~Y.} \bibnamefont{Chang}} \bibnamefont{and}
  \bibinfo{author}{\bibfnamefont{M.}~\bibnamefont{Boninsegni}},
  \bibinfo{journal}{J. Chem. Phys.} \textbf{\bibinfo{volume}{115}},
  \bibinfo{pages}{2629} (\bibinfo{year}{2001}).

\bibitem[{\citenamefont{Cuccoli et~al.}(1993)\citenamefont{Cuccoli, Macchi,
  Tognetti, and Vaia}}]{cu93}
\bibinfo{author}{\bibfnamefont{A.}~\bibnamefont{Cuccoli}},
  \bibinfo{author}{\bibfnamefont{A.}~\bibnamefont{Macchi}},
  \bibinfo{author}{\bibfnamefont{V.}~\bibnamefont{Tognetti}}, \bibnamefont{and}
  \bibinfo{author}{\bibfnamefont{R.}~\bibnamefont{Vaia}},
  \bibinfo{journal}{Phys. Rev. B} \textbf{\bibinfo{volume}{47}},
  \bibinfo{pages}{14923} (\bibinfo{year}{1993}).

\bibitem[{\citenamefont{M\"user et~al.}(1995)\citenamefont{M\"user, Nielaba,
  and Binder}}]{mu95}
\bibinfo{author}{\bibfnamefont{M.~H.} \bibnamefont{M\"user}},
  \bibinfo{author}{\bibfnamefont{P.}~\bibnamefont{Nielaba}}, \bibnamefont{and}
  \bibinfo{author}{\bibfnamefont{K.}~\bibnamefont{Binder}},
  \bibinfo{journal}{Phys. Rev. B} \textbf{\bibinfo{volume}{51}},
  \bibinfo{pages}{2723} (\bibinfo{year}{1995}).

\bibitem[{\citenamefont{Chakravarty}(2002)}]{ch02}
\bibinfo{author}{\bibfnamefont{C.}~\bibnamefont{Chakravarty}},
  \bibinfo{journal}{J. Chem. Phys.} \textbf{\bibinfo{volume}{116}},
  \bibinfo{pages}{8938} (\bibinfo{year}{2002}).

\bibitem[{\citenamefont{Neumann and Zoppi}(2002)}]{ne02}
\bibinfo{author}{\bibfnamefont{M.}~\bibnamefont{Neumann}} \bibnamefont{and}
  \bibinfo{author}{\bibfnamefont{M.}~\bibnamefont{Zoppi}},
  \bibinfo{journal}{Phys. Rev. E} \textbf{\bibinfo{volume}{65}},
  \bibinfo{pages}{031203} (\bibinfo{year}{2002}).

\bibitem[{\citenamefont{Herrero}(2002)}]{he02}
\bibinfo{author}{\bibfnamefont{C.~P.} \bibnamefont{Herrero}},
  \bibinfo{journal}{Phys. Rev. B} \textbf{\bibinfo{volume}{65}},
  \bibinfo{pages}{014112} (\bibinfo{year}{2002}).

\bibitem[{\citenamefont{Draeger and Ceperley}(2000)}]{dr00}
\bibinfo{author}{\bibfnamefont{E.~W.} \bibnamefont{Draeger}} \bibnamefont{and}
  \bibinfo{author}{\bibfnamefont{D.~M.} \bibnamefont{Ceperley}},
  \bibinfo{journal}{Phys. Rev. B} \textbf{\bibinfo{volume}{61}},
  \bibinfo{pages}{12094} (\bibinfo{year}{2000}).

\bibitem[{\citenamefont{Arms et~al.}(2003)\citenamefont{Arms, Shah, and
  Simmons}}]{ar03}
\bibinfo{author}{\bibfnamefont{D.~A.} \bibnamefont{Arms}},
  \bibinfo{author}{\bibfnamefont{R.~S.} \bibnamefont{Shah}}, \bibnamefont{and}
  \bibinfo{author}{\bibfnamefont{R.~O.} \bibnamefont{Simmons}},
  \bibinfo{journal}{Phys. Rev. B} \textbf{\bibinfo{volume}{67}},
  \bibinfo{pages}{094303} (\bibinfo{year}{2003}).

\bibitem[{\citenamefont{Venkataraman and Simmons}(2003)}]{ve03}
\bibinfo{author}{\bibfnamefont{C.~T.} \bibnamefont{Venkataraman}}
  \bibnamefont{and} \bibinfo{author}{\bibfnamefont{R.~O.}
  \bibnamefont{Simmons}}, \bibinfo{journal}{Phys. Rev. B}
  \textbf{\bibinfo{volume}{68}}, \bibinfo{pages}{224303}
  (\bibinfo{year}{2003}).

\bibitem[{\citenamefont{Moroni et~al.}(2000)\citenamefont{Moroni, Pederiva,
  Fantoni, and Boninsegni}}]{mo00}
\bibinfo{author}{\bibfnamefont{S.}~\bibnamefont{Moroni}},
  \bibinfo{author}{\bibfnamefont{F.}~\bibnamefont{Pederiva}},
  \bibinfo{author}{\bibfnamefont{S.}~\bibnamefont{Fantoni}}, \bibnamefont{and}
  \bibinfo{author}{\bibfnamefont{M.}~\bibnamefont{Boninsegni}},
  \bibinfo{journal}{Phys. Rev. Lett.} \textbf{\bibinfo{volume}{84}},
  \bibinfo{pages}{2650} (\bibinfo{year}{2000}).

\bibitem[{\citenamefont{Aziz et~al.}(1995)\citenamefont{Aziz, Janzen, and
  Moldover}}]{az95}
\bibinfo{author}{\bibfnamefont{R.~A.} \bibnamefont{Aziz}},
  \bibinfo{author}{\bibfnamefont{A.~R.} \bibnamefont{Janzen}},
  \bibnamefont{and} \bibinfo{author}{\bibfnamefont{M.~R.}
  \bibnamefont{Moldover}}, \bibinfo{journal}{Phys. Rev. Lett.}
  \textbf{\bibinfo{volume}{74}}, \bibinfo{pages}{1586} (\bibinfo{year}{1995}).

\bibitem[{\citenamefont{Bruch and McGee}(1973)}]{br73}
\bibinfo{author}{\bibfnamefont{L.~W.} \bibnamefont{Bruch}} \bibnamefont{and}
  \bibinfo{author}{\bibfnamefont{I.~J.} \bibnamefont{McGee}},
  \bibinfo{journal}{J. Chem. Phys.} \textbf{\bibinfo{volume}{59}},
  \bibinfo{pages}{409} (\bibinfo{year}{1973}).

\bibitem[{\citenamefont{Loubeyre}(1987)}]{lo87}
\bibinfo{author}{\bibfnamefont{P.}~\bibnamefont{Loubeyre}},
  \bibinfo{journal}{Phys. Rev. Lett.} \textbf{\bibinfo{volume}{58}},
  \bibinfo{pages}{1857} (\bibinfo{year}{1987}).

\bibitem[{\citenamefont{Gillan}(1988)}]{gi88}
\bibinfo{author}{\bibfnamefont{M.~J.} \bibnamefont{Gillan}},
  \bibinfo{journal}{Phil. Mag. A} \textbf{\bibinfo{volume}{58}},
  \bibinfo{pages}{257} (\bibinfo{year}{1988}).

\bibitem[{\citenamefont{Noya et~al.}(1996)\citenamefont{Noya, Herrero, and
  Ram\'{\i}rez}}]{no96}
\bibinfo{author}{\bibfnamefont{J.~C.} \bibnamefont{Noya}},
  \bibinfo{author}{\bibfnamefont{C.~P.} \bibnamefont{Herrero}},
  \bibnamefont{and}
  \bibinfo{author}{\bibfnamefont{R.}~\bibnamefont{Ram\'{\i}rez}},
  \bibinfo{journal}{Phys. Rev. B} \textbf{\bibinfo{volume}{53}},
  \bibinfo{pages}{9869} (\bibinfo{year}{1996}).

\bibitem[{\citenamefont{Chandler and Wolynes}(1981)}]{ch81}
\bibinfo{author}{\bibfnamefont{D.}~\bibnamefont{Chandler}} \bibnamefont{and}
  \bibinfo{author}{\bibfnamefont{P.~G.} \bibnamefont{Wolynes}},
  \bibinfo{journal}{J. Chem. Phys.} \textbf{\bibinfo{volume}{74}},
  \bibinfo{pages}{4078} (\bibinfo{year}{1981}).

\bibitem[{\citenamefont{Singer and Smith}(1988)}]{si88}
\bibinfo{author}{\bibfnamefont{K.}~\bibnamefont{Singer}} \bibnamefont{and}
  \bibinfo{author}{\bibfnamefont{W.}~\bibnamefont{Smith}},
  \bibinfo{journal}{Mol. Phys.} \textbf{\bibinfo{volume}{64}},
  \bibinfo{pages}{1215} (\bibinfo{year}{1988}).

\bibitem[{\citenamefont{Cuccoli et~al.}(1997)\citenamefont{Cuccoli, Macchi,
  Pedrolli, Tognetti, and Vaia}}]{cu97}
\bibinfo{author}{\bibfnamefont{A.}~\bibnamefont{Cuccoli}},
  \bibinfo{author}{\bibfnamefont{A.}~\bibnamefont{Macchi}},
  \bibinfo{author}{\bibfnamefont{G.}~\bibnamefont{Pedrolli}},
  \bibinfo{author}{\bibfnamefont{V.}~\bibnamefont{Tognetti}}, \bibnamefont{and}
  \bibinfo{author}{\bibfnamefont{R.}~\bibnamefont{Vaia}},
  \bibinfo{journal}{Phys. Rev. B} \textbf{\bibinfo{volume}{56}},
  \bibinfo{pages}{51} (\bibinfo{year}{1997}).

\bibitem[{\citenamefont{Noya et~al.}()\citenamefont{Noya, Herrero, and
  Ram\'{\i}rez}}]{no96b}
\bibinfo{author}{\bibfnamefont{J.~C.} \bibnamefont{Noya}},
  \bibinfo{author}{\bibfnamefont{C.~P.} \bibnamefont{Herrero}},
  \bibnamefont{and}
  \bibinfo{author}{\bibfnamefont{R.}~\bibnamefont{Ram\'{\i}rez}},
  \bibinfo{note}{unpublished}.

\bibitem[{\citenamefont{Herrero and Ram\'{\i}rez}(2001)}]{he01}
\bibinfo{author}{\bibfnamefont{C.~P.} \bibnamefont{Herrero}} \bibnamefont{and}
  \bibinfo{author}{\bibfnamefont{R.}~\bibnamefont{Ram\'{\i}rez}},
  \bibinfo{journal}{Phys. Rev. B} \textbf{\bibinfo{volume}{63}},
  \bibinfo{pages}{024103} (\bibinfo{year}{2001}).

\bibitem[{\citenamefont{Herrero and Ram\'{\i}rez}(2005)}]{he05}
\bibinfo{author}{\bibfnamefont{C.~P.} \bibnamefont{Herrero}} \bibnamefont{and}
  \bibinfo{author}{\bibfnamefont{R.}~\bibnamefont{Ram\'{\i}rez}},
  \bibinfo{journal}{Phys. Rev. B} \textbf{\bibinfo{volume}{71}},
  \bibinfo{pages}{174111} (\bibinfo{year}{2005}).

\bibitem[{\citenamefont{Herrero}(2003)}]{he03a}
\bibinfo{author}{\bibfnamefont{C.~P.} \bibnamefont{Herrero}},
  \bibinfo{journal}{J. Phys.: Condens. Matter} \textbf{\bibinfo{volume}{15}},
  \bibinfo{pages}{475} (\bibinfo{year}{2003}).

\bibitem[{\citenamefont{Karasevskii and Holzapfel}(2003)}]{ka03}
\bibinfo{author}{\bibfnamefont{A.~I.} \bibnamefont{Karasevskii}}
  \bibnamefont{and} \bibinfo{author}{\bibfnamefont{W.~B.}
  \bibnamefont{Holzapfel}}, \bibinfo{journal}{Phys. Rev. B}
  \textbf{\bibinfo{volume}{67}}, \bibinfo{pages}{224301}
  (\bibinfo{year}{2003}).

\bibitem[{\citenamefont{Lawler et~al.}(2004)\citenamefont{Lawler, Chang, and
  Shirley}}]{la04}
\bibinfo{author}{\bibfnamefont{H.~M.} \bibnamefont{Lawler}},
  \bibinfo{author}{\bibfnamefont{E.~K.} \bibnamefont{Chang}}, \bibnamefont{and}
  \bibinfo{author}{\bibfnamefont{E.~L.} \bibnamefont{Shirley}},
  \bibinfo{journal}{Phys. Rev. B} \textbf{\bibinfo{volume}{69}},
  \bibinfo{pages}{174104} (\bibinfo{year}{2004}).

\bibitem[{\citenamefont{Pollock et~al.}(1972)\citenamefont{Pollock, Bruce,
  Chester, and Krumhansl}}]{po72}
\bibinfo{author}{\bibfnamefont{E.~L.} \bibnamefont{Pollock}},
  \bibinfo{author}{\bibfnamefont{T.~A.} \bibnamefont{Bruce}},
  \bibinfo{author}{\bibfnamefont{G.~V.} \bibnamefont{Chester}},
  \bibnamefont{and} \bibinfo{author}{\bibfnamefont{J.~A.}
  \bibnamefont{Krumhansl}}, \bibinfo{journal}{Phys. Rev. B}
  \textbf{\bibinfo{volume}{5}}, \bibinfo{pages}{4180} (\bibinfo{year}{1972}).

\bibitem[{\citenamefont{Holian et~al.}(1973)\citenamefont{Holian, Gwinn, Luntz,
  and Alder}}]{ho73}
\bibinfo{author}{\bibfnamefont{B.~L.} \bibnamefont{Holian}},
  \bibinfo{author}{\bibfnamefont{W.~D.} \bibnamefont{Gwinn}},
  \bibinfo{author}{\bibfnamefont{A.~C.} \bibnamefont{Luntz}}, \bibnamefont{and}
  \bibinfo{author}{\bibfnamefont{B.~J.} \bibnamefont{Alder}},
  \bibinfo{journal}{J. Chem. Phys.} \textbf{\bibinfo{volume}{59}},
  \bibinfo{pages}{5444} (\bibinfo{year}{1973}).

\bibitem[{\citenamefont{Etters and Danilowicz}(1974)}]{et74}
\bibinfo{author}{\bibfnamefont{R.~D.} \bibnamefont{Etters}} \bibnamefont{and}
  \bibinfo{author}{\bibfnamefont{R.~L.} \bibnamefont{Danilowicz}},
  \bibinfo{journal}{Phys. Rev. A} \textbf{\bibinfo{volume}{9}},
  \bibinfo{pages}{1698} (\bibinfo{year}{1974}).

\end{thebibliography}

\end{document}